\documentclass[prd,superscriptaddress,amsfonts,amssymb,amsmath,showpacs,twocolumn]{revtex4-2}
\usepackage{bm}
\usepackage{amsfonts}
\usepackage{latexsym}
\usepackage[latin1]{inputenc}
\usepackage{graphicx}
\usepackage{amsmath}
\usepackage{palatino}
\usepackage{ragged2e}
\usepackage{mathpazo}
\usepackage{textcomp}
\usepackage{float}
\usepackage{booktabs}
\usepackage{dcolumn}
\usepackage{multirow}
\usepackage{hyperref}
\hypersetup{colorlinks,citecolor=blue}
\usepackage{amsmath}
\usepackage{xcolor}
\usepackage{orcidlink}
\usepackage[caption=false]{subfig}
\usepackage{commath}
\def\jnl@style{\it}
\def\aaref@jnl#1{{\jnl@style#1}}

\def\aaref@jnl#1{{\jnl@style#1}}

\def\aj{\aaref@jnl{AJ}}                   
\def\apj{\aaref@jnl{ApJ}}                 
\def\apjl{\aaref@jnl{ApJ}}                
\def\apjs{\aaref@jnl{ApJS}}               
\def\apss{\aaref@jnl{Ap\&SS}}             
\def\aap{\aaref@jnl{A\&A}}                
\def\aapr{\aaref@jnl{A\&A~Rev.}}          
\def\aaps{\aaref@jnl{A\&AS}}              
\def\mnras{\aaref@jnl{Mon.~Not.~Roy.~Astron.~Soc.}}             
\def\prd{\aaref@jnl{Phys.~Rev.~D}}        
\def\prc{\aaref@jnl{Phys.~Rev.~C}}  
\def\prl{\aaref@jnl{Phys.~Rev.~Lett.}}    
\def\qjras{\aaref@jnl{QJRAS}}             
\def\skytel{\aaref@jnl{S\&T}}             
\def\ssr{\aaref@jnl{Space~Sci.~Rev.}}     
\def\zap{\aaref@jnl{ZAp}}                 
\def\nat{\aaref@jnl{Nature}}              
\def\aplett{\aaref@jnl{Astrophys.~Lett.}} 
\def\apspr{\aaref@jnl{Astrophys.~Space~Phys.~Res.}} 
\def\physrep{\aaref@jnl{Phys.~Rep.}}      
\def\physscr{\aaref@jnl{Phys.~Scr}}       
\def\commat{\aaref@jnl{Comm.~Math.~Phys.}}              
\def\science{\aaref@jnl{Science}}               
\def\cqg{\aaref@jnl{Classical Quant.~Grav.}}            
\def\jpcs{\aaref@jnl{JPCS}}                                     
\def\ijmpd{\aaref@jnl{Int.~J.~Mod.~Phys.~D}}                    
\def\grg{\aaref@jnl{Gen.~Relat.~Gravit.}}               
\def\rpp{\aaref@jnl{Rep.~Prog.~Phys.}}          
\def\npa{\aaref@jnl{Nucl.~Phys.~A}}        
\def\lrr{\aaref@jnl{Living Rev.~Rel.}}                   
\def\jcap{\aaref@jnl{J.~Cosmology Astropart.~Phys.}}    
\def\rmp{\aaref@jnl{Rev.~Mod.~Phys.}}   
\def\epjc{\aaref@jnl{Eur.~Phys.~J.~C}}

\allowdisplaybreaks[1]

\begin{document}

\color{black}       

\title{Constraining viscous dark energy equation of state in $f(R,L_m)$ gravity}

\author{Lakhan V. Jaybhaye\orcidlink{0000-0003-1497-276X}}
\email{lakhanjaybhaye@gmail.com}
\affiliation{Department of Mathematics, Birla Institute of Technology and
Science-Pilani,\\ Hyderabad Campus, Hyderabad-500078, India.}
\author{Raja Solanki\orcidlink{0000-0001-8849-7688}}
\email{rajasolanki8268@gmail.com}
\affiliation{Department of Mathematics, Birla Institute of Technology and
Science-Pilani,\\ Hyderabad Campus, Hyderabad-500078, India.}
\author{Sanjay Mandal\orcidlink{0000-0003-2570-2335}}
\email{sanjaymandal960@gmail.com}
\affiliation{Department of Mathematics, Birla Institute of Technology and
Science-Pilani,\\ Hyderabad Campus, Hyderabad-500078, India.}
\author{P.K. Sahoo\orcidlink{0000-0003-2130-8832}}
\email{pksahoo@hyderabad.bits-pilani.ac.in}
\affiliation{Department of Mathematics, Birla Institute of Technology and
Science-Pilani,\\ Hyderabad Campus, Hyderabad-500078, India.}

\date{\today}

\begin{abstract}

In this article, we attempt to describe cosmic late-time acceleration of the universe in the framework of $f(R,L_m)$ gravity by using an effective equation of state when the account is taken of bulk viscosity. We presume a non-linear $f(R,L_m)$ functional form, specifically, $f(R,L_m)=\frac{R}{2}+L_m^\alpha $, where $\alpha$ is free model parameter. We obtain the exact solution of our bulk viscous matter dominated $f(R,L_m)$ model, and then we utilize the combined $H(z)+Pantheon+Analysis$ data sets to estimate the best fit values of the free parameters of our model. Then we characterize the behaviour of the matter-energy density, effective pressure, and the equation of state (EoS) parameter incorporating the viscous type fluid. The evolution profile of the effective EoS parameter depicts an acceleration phase of the cosmic expansion whereas the pressure with the effect of viscosity exhibits negative behaviour that can lead to the accelerating expansion of the universe. Moreover, the cosmic matter-energy density shows expected positive behaviour. Further, we investigate the behaviour of statefinder parameters for the assumed $f(R,L_m)$ model. We find that the evolutionary trajectory of the given model lies in the quintessence region. In addition, we employ the Om diagnostic test that indicates our model exhibits quintessence behavior. Lastly, we check the energy condition criteria and find that violation of SEC occurs in the past, whereas NEC and DEC satisfies the positivity criteria. We find that our $f(R,L_m)$ cosmological model with the effect of bulk viscosity provides a good fit of the recent observational data and can efficiently describe the cosmic expansion scenario.

\end{abstract}

\maketitle

\section{Introduction}\label{sec1}
\justify 

Cosmology has faced a dramatic change when the observational evidence from type Ia supernovae searches \cite{Riess,Perlmutter} confirmed the accelerating behaviour of the cosmic expansion. In the last two decades, a plethora of observational results such as Large Scale Structure \cite{E11}, Wilkinson Microwave Anisotropy Probe \cite{D.N.}, Cosmic Microwave Background Radiation \cite{C.L.,R.R.}, and the Baryonic Acoustic Oscillations \cite{D.J.,W.J.} agrees with the observed cosmic acceleration. The prominent explanation to describe this accelerating scenario is the presence of a dark energy component characterized by an equation of state $\omega=-1.018 \pm 0.057$ for a flat universe \cite{Lake}. Another promising way to describe the accelerating expansion of the universe by bypassing the undetected dark energy component is to consider that the more generic action describes the gravitational field. The cosmological models in which the Einstein-Hilbert action of general relativity (GR) is modified by introducing the generic function $f(R)$, where $R$ denotes the Ricci scalar curvature, first proposed in \cite{H.A.,R.K.,H.K.}. The $f(R)$ gravity model is capable to describe the expansion mechanism without invoking any exotic dark energy component \cite{Carr,Cap}. Observational signatures of $f(R)$ gravity models along with the solar system and equivalence principle constraints, presented in the references \cite{Shin,Sal,Sean,Alex}. In the context of solar system tests, the viable cosmological models of $f(R)$ gravity do exists \cite{Noj,V.F.,L.A.}. Odintsov et al. have analysed the $H_0$ tension and the role of energy conditions in $f(R)$ gravity models \cite{Odi-1,Odi-2}. One can follow the references \cite{Noj-2,Noj-3,JS,Noj-4,Odi-3,Odi-4,AP} to see the various implications of cosmological models of $f(R)$ gravity.

A generalization of the curvature based $f(R)$ gravity that incorporates an explicit coupling of the generic function $f(R)$ with the matter Lagrangian density $L_m$ appeared in \cite{O.B.}. This coupling case was further extended to the case of arbitrary matter geometry couplings \cite{THK}. Harko and Lobo investigated the curvature-matter couplings in modified gravity from linear aspects to conformally invariant theories \cite{L-Ex}. Models with non-minimal matter geometry couplings have great astrophysical and cosmological implications. Harko studied the galactic rotation curves, the matter Lagrangian and the energy momentum tensor, thermodynamical features, coupling matter and curvature in Weyl geometry, in the context of non-minimal couplings \cite{THK-2,THK-3,THK-4,THK-5}. Moreover, Bertolami et al. \cite{THK-EX} investigated curvature-matter couplings in modified gravity and Faraoni examined the viability criterion for modified gravity with an extra force \cite{V.F.-2}.  Further, Harko and Lobo recently proposed \cite{THK-6} $f(R,L_m)$ gravity theory that is a generalization matter curvature coupling theories, where $f(R,L_m)$ is a generic function that depends on the Ricci scalar $R$ and the matter Lagrangian $L_m$. In this theory, the covariant divergence of the stress-energy tensor does not vanishes, an extra force orthogonal to four velocities arises, and the motion of test particle is non-geodesic. The models of $f(R,L_m)$ gravity theory disobey the equivalence principle, and that is restrained by the solar system experimental tests \cite{FR,JP}. Recently, several interesting results on $f(R,L_m)$ gravity have been appeared, for instance, see references \cite{GM,RV-1,RV-2,Jay}. 

\justify In the presented manuscript, we are going to explore the cosmological $f(R,L_m)$ model that exhibits viscous type fluid. The introduction of the coefficient of viscosity in the models of cosmology has a long history. From a hydrodynamicist's point of view, there are two viscosity coefficients commonly appeared in the literature, namely the bulk viscosity coefficient $\zeta$ and the shear viscosity coefficient $\eta$. By assuming the observationally supported spatial isotropy of the cosmos, shear viscosity can be omitted. Whenever a system gets deviated from its thermal equilibrium, then to recover its thermal equilibrium state an effective pressure is generated. Bulk viscosity in a cosmological fluid is the manifestation of such an effective pressure. The idea is to consider the bulk viscosity coefficient $\zeta$ in the $f(R,L_m)$ gravity model. We assume that the coefficient of bulk viscosity $\zeta$ satisfies a scaling law and that reduces the Einstein case to a form proportional to the Hubble parameter. It has been appeared that this scaling law is quite useful. One can check the references to review some interesting viscous fluid cosmological models \cite{IB-1,IB-2,IB-3,IB-4,IB-5,JM,AVS,MAT}.

The present manuscript is organized in the following manner. In Sec \ref{sec2}, we present the action and basic formulation governing the dynamics in $f(R,L_m)$ gravity. In Sec \ref{sec3}, we present the Friedmann like equations corresponding to the flat FLRW universe. In Sec \ref{sec4}, we assume a $f(R,L_m)$ functional and then we calculate the expression for the Hubble parameter and the equation of state (EoS) parameter relating the pressure term of bulk viscous matter with its energy density. In section Sec \ref{sec5}, we estimate the values of the $H_0$ and model parameters that obeys with observations, by incorporating the combined H(z)+Pantheon+Analysis data sets. In addition, we characterize the behavior of various parameters such as density, effective pressure, and the EoS parameter. Further in Sec \ref{sec6}, we investigate the $r-s$ parameter trajectory of our $f(R,L_m)$ model to check the dark energy behavior recognized by the assumed model. Moreover, in sec \ref{sec7} and sec \ref{sec8}, we employ the Om diagnostic test and energy condition criteria. Finally, in Sec \ref{sec9}, we conclude our findings. 

\section{ $f(R,L_m)$ Gravity Theory}\label{sec2}

\justify

The generic action for $f(R,L_m)$ gravity read as

\begin{equation}\label{1a}
S= \int{f(R,L_m)\sqrt{-g}d^4x} 
\end{equation}
Here $R$ represents the Ricci scalar curvature and $L_m$ denotes the matter Lagrangian . 

One can obtained the Ricci scalar $R$ by contracting the Ricci tensor $R_{\mu\nu}$ as
\begin{equation}\label{1b}
R= g^{\mu\nu} R_{\mu\nu}
\end{equation} 
where the Ricci tensor is given by
 
\begin{equation}\label{1c}
R_{\mu\nu}= \partial_\lambda \Gamma^\lambda_{\mu\nu} - \partial_\mu \Gamma^\lambda_{\lambda\nu} + \Gamma^\lambda_{\mu\nu} \Gamma^\sigma_{\sigma\lambda} - \Gamma^\lambda_{\nu\sigma} \Gamma^\sigma_{\mu\lambda}
\end{equation}
with $\Gamma^\alpha_{\beta\gamma}$ representing the components of Levi-Civita connection.


Now we obtained the following field equation governing the dynamics of gravitational interactions, by varying the action \eqref{1a} with respect to the metric tensor $g_{\mu\nu}$,

\begin{equation}\label{1d}
f_R R_{\mu\nu} + (g_{\mu\nu} \square - \nabla_\mu \nabla_\nu)f_R - \frac{1}{2} (f-f_{L_m}L_m)g_{\mu\nu} = \frac{1}{2} f_{L_m} T_{\mu\nu}
\end{equation}

Here $f_R \equiv \frac{\partial f}{\partial R}$, $f_{L_m} \equiv \frac{\partial f}{\partial L_m}$, and $T_{\mu\nu}$ represents the stress-energy tensor for the cosmic fluid, defined by 

\begin{equation}\label{1e}
T_{\mu\nu} = \frac{-2}{\sqrt{-g}} \frac{\delta(\sqrt{-g}L_m)}{\delta g^{\mu\nu}}
\end{equation}

The connection among the energy-momentum scalar $T$, the matter Lagrangian term $L_m$, and the Ricci scalar curvature $R$ acquired by contracting the field equation \eqref{1d} as

\begin{equation}\label{1f}
R f_R + 3\square f_R - 2(f-f_{L_m}L_m) = \frac{1}{2} f_{L_m} T
\end{equation}

Here $\square F = \frac{1}{\sqrt{-g}} \partial_\alpha (\sqrt{-g} g^{\alpha\beta} \partial_\beta F)$ for any scalar function $F$ .

In addition, one can obtain the following relation by employing the covariant derivative in equation \eqref{1d}

\begin{equation}\label{1g}
\nabla^\mu T_{\mu\nu} = 2\nabla^\mu ln(f_{L_m}) \frac{\partial L_m}{\partial g^{\mu\nu}}
\end{equation} \\

\section{Motion equations in $f(R,L_m)$ gravity}\label{sec3}

\justify

In order to probe the cosmological implications, we consider the following homogeneous and spatially isotropic FLRW metric \cite{Ryden}

\begin{equation}\label{2a}
ds^2= -dt^2 + a^2(t)[dx^2+dy^2+dz^2]
\end{equation}

\justify where, $ a(t) $ is the cosmic scale factor.
The Ricci scalar obtained for the metric \eqref{2a} is

\begin{equation}\label{2b}
R= 6 \frac{\ddot{a}}{a}+ 6 \bigl( \frac{\dot{a}}{a} \bigr)^2 = 6 ( \dot{H}+2H^2 )
\end{equation}
where $H=\frac{\dot{a}}{a}$ is the Hubble parameter.

\justify The energy-momentum tensor comprises of energy density $\rho$ and the pressure $\bar{p}$ of the cosmic fluid with  viscosity effect is given by,

\begin{equation}\label{2c}
\mathcal{T}_{\mu\nu}=(\rho+\bar{p})u_\mu u_\nu + \bar{p}g_{\mu\nu}
\end{equation}

where $\bar{p}=p-3\zeta H$ and $u^\mu=(1,0,0,0)$ are components of the four velocities. Here $p$ is the usual pressure and $\zeta > 0$ is the coefficient of bulk viscosity.

\justify The connection between matter-energy density and the usual pressure is given as \cite{J}
\begin{equation}\label{2d}
p=(\gamma-1)\rho
\end{equation}
where $\gamma$ is a constant with $0 \leq \gamma \leq 2$. Hence the effective equation of state characterizing the bulk viscous cosmic fluid reads as \cite{brevik/2005,gron/1990,C.E./1940}
\begin{equation}\label{2e}
\bar{p}= (\gamma-1)\rho -3\zeta H
\end{equation}

\justify Under the constraint of homogeneity and spatial isotropy, the cosmic fluid incorporating viscosity possesses dissipative phenomenon. Considering viscosity in a cosmic fluid can minimize the ideal characteristics of a fluid, and participates to the total pressure negatively. This can be checked in References \cite{odintsov/2020,fabris/2006,meng/2009}. 

\justify The Friedmann equations that characterizes the bulk viscous matter dominated universe in $f(R,L_m)$ gravity reads as \cite{LK}

\begin{equation}\label{2f}
3H^2 f_R + \frac{1}{2} \left( f-f_R R-f_{L_m}L_m \right) + 3H \dot{f_R}= \frac{1}{2}f_{L_m} \rho 
\end{equation}
and
\begin{equation}\label{2g}
\dot{H}f_R + 3H^2 f_R - \ddot{f_R} -3H\dot{f_R} + \frac{1}{2} \left( f_{L_m}L_m - f \right) = \frac{1}{2} f_{L_m}\bar{p}
\end{equation}

\section{Cosmological $f(R,L_m)$ Model }\label{sec4}

We choose the following $f(R,L_m)$ function in order to explore the dynamics of the universe possesses viscosity \cite{LK,LB},

\begin{equation}\label{3a} 
f(R,L_m)=\frac{R}{2}+L_m^\alpha 
\end{equation}

Here $\alpha$ is free model parameter. The model under consideration is more general in nature and it is motivated by the
functional form $f(R,L_m) = f_1(R)+f_2(R) G(L_m)$ that represents arbitrary matter-geometry coupling \cite{LB}. 

\justify Then for this specific functional form with $L_m=\rho$ \cite{HLR}, the Friedmann equations \eqref{2f} and \eqref{2g} characterizing the universe dominated with bulk viscous matter becomes

\begin{equation}\label{3b}
3H^2=(2\alpha-1) \rho^\alpha
\end{equation}

and

\begin{equation}\label{3c}
2\dot{H}+3H^2=  \left\{ (\alpha-1)\rho-\alpha \bar{p} \right\} \rho^{\alpha-1} 
\end{equation}

Now by using equation \eqref{1g}, we obtained the following matter conservation equation for our bulk viscous cosmological $f(R,L_m)$ model

\begin{equation}\label{3d}
(2\alpha-1)\dot{\rho}+ 3\gamma H \rho = 0
\end{equation}

From equations \eqref{3b} and \eqref{3c}, one can have

\begin{equation}\label{3e}
\dot{H}+\frac{3\alpha\gamma}{2(2\alpha-1)} H^2 = \frac{3}{2} \left( \frac{3}{2\alpha-1} \right)^{\frac{\alpha-1}{\alpha}}\alpha\zeta H^{\frac{3\alpha-2}{\alpha}}
\end{equation}

We substitute $ \frac{1}{H} \frac{d}{dt}= \frac{d}{dln(a)}$ so that equation \eqref{3e} becomes

\begin{equation}\label{3f}
\frac{dH}{dln(a)}+\frac{3\alpha\gamma}{2(2\alpha-1)} H = \frac{3}{2} \left( \frac{3}{2\alpha-1} \right)^{\frac{\alpha-1}{\alpha}}\alpha\zeta H^{\frac{2(\alpha-1)}{\alpha}}
\end{equation}

On integrating the equation \eqref{3f} we obtained the expression for Hubble parameter as follows

\begin{widetext}
\begin{equation}\label{3g}
H(z)=\big\{ H_0^{\frac{2-\alpha}{\alpha}} (1+z)^{\frac{3\gamma(2-\alpha)}{2(2\alpha-1)}} + \frac{3\zeta}{\gamma} \left( \frac{2\alpha-1}{3} \right)^{\frac{1}{\alpha}} [ 1-(1+z)^{\frac{3\gamma(2-\alpha)}{2(2\alpha-1)}} ] \big\}^\frac{\alpha}{2-\alpha}
\end{equation}
\end{widetext}

where $H(0)=H_0$ represents the present value of the Hubble parameter. In particular, for the case $\alpha=1$ with $\gamma=1$ and $\zeta=0$, the solution reduces to $H(z)=H_0(1+z)^{\frac{3}{2}}$ , the usual ordinary matter dominated universe. 

\justify The effective equation of state parameter for our bulk viscous cosmological model is given by
\begin{equation}\label{3h}
\omega_{eff} = \frac{p_{eff}}{\rho} = \gamma -1 - \frac{3\zeta H}{\rho}
\end{equation}
 
By using equations \eqref{3b} and \eqref{3g}, one can acquired
\begin{widetext}
\begin{equation}\label{3i}
\omega_{eff} = \gamma -1 - 3\zeta \left( \frac{2\alpha-1}{3} \right)^{\frac{1}{\alpha}} \big\{ H_0^{\frac{2-\alpha}{\alpha}} (1+z)^{\frac{3\gamma(2-\alpha)}{2(2\alpha-1)}} + \frac{3\zeta}{\gamma} \left( \frac{2\alpha-1}{3} \right)^{\frac{1}{\alpha}} [ 1-(1+z)^{\frac{3\gamma(2-\alpha)}{2(2\alpha-1)}} ] \big\}^{-1}
\end{equation}
\end{widetext}

\section{Data, Methodology, and Physical Interpretation}\label{sec5}

In this section, we estimate the parameter values of our model that is appropriate to describe the various cosmic epochs, by invoking the $H(z)$ and Pantheon+Analysis data sets. To calculate the suitable values of $H_0$ and model parameters $\alpha$, $\gamma$, and $\zeta$, we incorporate $31$ points of $H(z)$ data sets and $1701$ points from the Pantheon+Analysis samples. To estimate the mean values of the parameters of our viscosity model, we apply the Bayesian technique and likelihood function along with the Markov Chain Monte Carlo (MCMC) method in \texttt{emcee} python library \cite{Mackey/2013}. 

\subsubsection{H(z) datasets}

It is well known that the Hubble parameter can directly investigate cosmic expansion. In terms of redshift the Hubble parameter can be acquired as $H(z)=-\frac{1}{1+z}\frac{dz}{dt}$. Since $dz$ is derived from the spectroscopic surveys therefore one can obtain the model-independent $H(z)$ value by measuring the $dt$. In this manuscript, we incorporate $31$ data points of $H(z)$ measurements in the redshift range $0.07 \leq z \leq 2.41$ \cite{GSS}. One can check the reference \cite{RS} for the complete list of $31$ data points. We define the chi-square function to find out the mean values of the bulk viscous model parameters $\alpha$, $\gamma$, $\zeta$, and $H_0$ as follows,

\begin{equation}\label{4a}
\chi _{H}^{2}(H_0,\alpha,\gamma,\zeta)=\sum\limits_{k=1}^{31}
\frac{[H_{th}(z_{k},H_0,\alpha,\gamma,\zeta)-H_{obs}(z_{k})]^{2}}{
\sigma _{H(z_{k})}^{2}}.  
\end{equation}

Here, the theoretical value of the $H(z)$ acquired by our cosmological  model is represented by $H_{th}$ whereas $H_{obs}$ denotes its observed value and $\sigma_{H(z_{k})}$ is the standard error. 

\subsubsection{Pantheon datasets}

Earlier, the observational results on type Ia supernovae confirmed that our universe is going through a phase of accelerated expansion. In the past two decades, observations on supernovae samples have been extensively increased. In 2018, $1048$ samples of type Ia supernovae covering the redshift range $0.01 < z < 2.3$ has been released which is known as Pantheon supernovae samples \cite{Scolnic/2018}. The PanSTARSS1 Medium, Deep Survey, SDSS, HST surveys, SNLS, and numerous low redshift surveys contribute to it. Recently, Pantheon+ Analysis sample incorporating 1701 light curves of $1550$ supernovae in the range of redshift $[0.001, 2.26]$ has been released \cite{Brout}. The luminosity distance is taken to be \cite{planck_collaboration/2020},

\begin{eqnarray*}
D_{L}(z) &=&\frac{c(1+z)}{H_{0}}S_{k}\left( H_{0}\int_{0}^{z}\frac{1}{%
H(z')}dz'\right) , \\
\text{where }S_{k}(x) &=&\left\{ 
\begin{array}{c}
\sinh (x\sqrt{\Omega _{k}})/\Omega _{k}\text{, }\Omega _{k}>0 \\ 
x\text{, \ \ \ \ \ \ \ \ \ \ \ \ \ \ \ \ \ \ \ \ \ \ \ }\Omega _{k}=0 \\ 
\sin x\sqrt{\left\vert \Omega _{k}\right\vert })/\left\vert \Omega
_{k}\right\vert \text{, }\Omega _{k}<0%
\end{array}%
\right. 
\end{eqnarray*}%

\justify For a spatially flat universe, we have

\begin{equation}\label{4b}
D_{L}(z)= (1+z) \int_{0}^{z} \frac{c dz'}{H(z')},
\end{equation}
where $c$ is the speed of light.

\justify We have calculated the $\chi^{2}$ function for the Pantheon supernovae samples by correlating the theoretical distance modulus 

\begin{equation}\label{4c}
\mu(z)= 5log_{10}D_{L}(z)+\mu_{0}, 
\end{equation}
with 
\begin{equation}\label{4d}
\mu_{0} =  5log(1/H_{0}Mpc) + 25,
\end{equation}
such that
\begin{equation}\label{4e}
\chi^2_{SN}(p_1,....)=\sum_{i,j=1}^{1701}\bigtriangledown\mu_{i}\left(C^{-1}_{SN}\right)_{ij}\bigtriangledown\mu_{j},
\end{equation}

\justify Here $p_j$ represents free model parameters and $C_{SN}$ is the covariance matrix \cite{Brout}, and
 \begin{align*}
  \quad \bigtriangledown\mu_{i}=\mu^{th}(z_i,p_1,...)-\mu_i^{obs}.
 \end{align*}
 
\justify where $\mu_{th}$ represents value of the distance modulus predicted by our model while $\mu_{obs}$  its observed value.

\justify Now the $\chi^{2}$ function for the H(z)+Pantheon+Analysis data sets is taken to be

\begin{equation}
\chi^{2}_{total}= \chi^{2}_H + \chi^{2}_{SN} 
\end{equation}

\justify We present the $1-\sigma$ and $2-\sigma$ likelihood contours for the model parameters $\alpha$, $\gamma$, $\zeta$, and $H_0$ using combined H(z)+Pantheon+Analysis data sets below.

\begin{widetext}

\begin{figure}[H]
\centering
\includegraphics[scale=0.9]{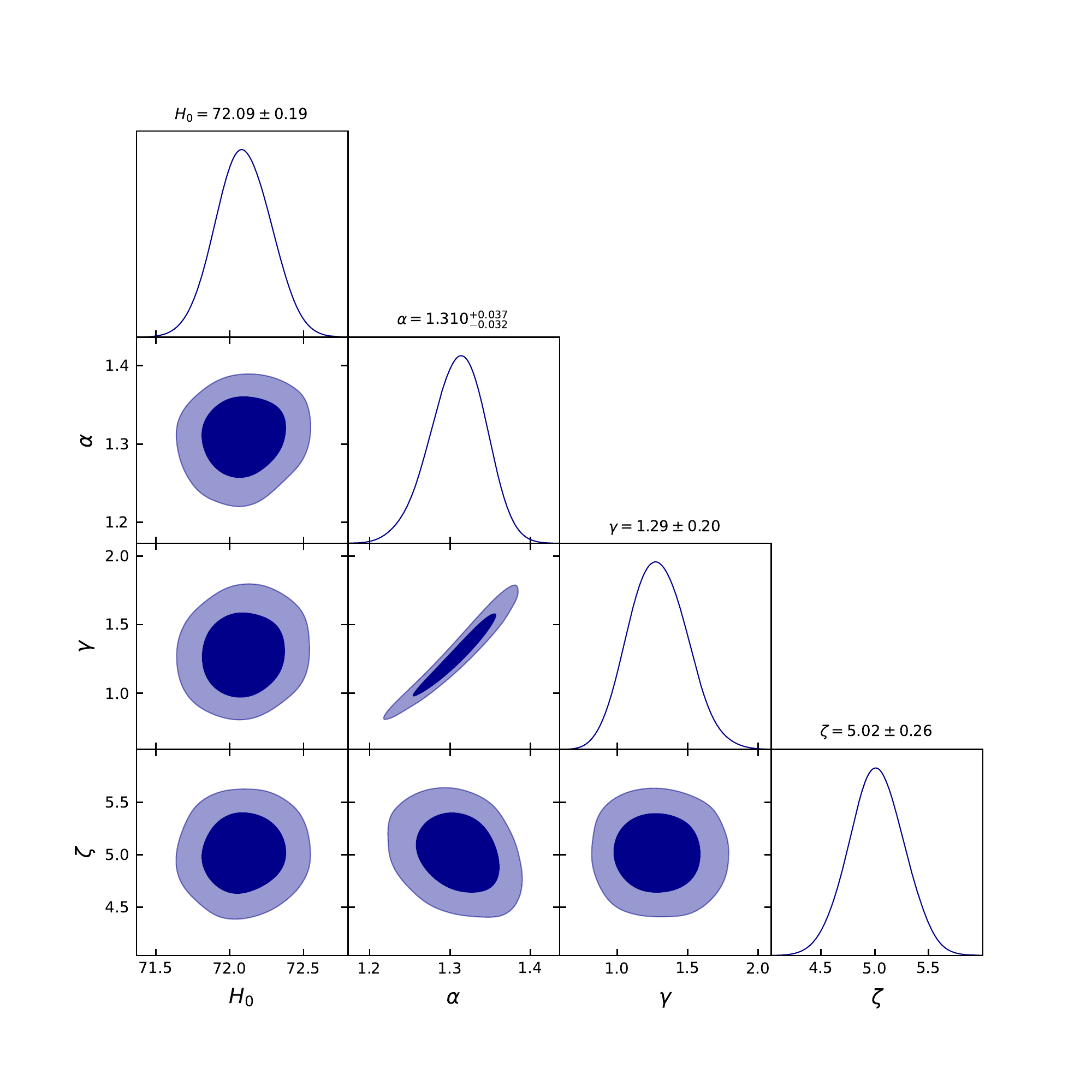}
\caption{The $1-\sigma$ and $2-\sigma$ contours for the model parameters $\alpha$, $\gamma$, $\zeta$, and $H_0$ using combined H(z)+Pantheon+Analysis data sets}\label{f1}
\end{figure}

\end{widetext}

\justify The obtained best fit values are $\alpha=1.310^{+0.037}_{-0.032}$, $\gamma = 1.29 \pm 0.20$, $\zeta= 5.02 \pm 0.26$, and $H_0= 72.09 \pm 0.19$. 

\justify Now we are going to present the cosmological implications of obtained observational constraints. We analyze the behaviour of energy density, pressure component incorporating viscosity, and the effective EoS parameter for the obtained mean values of $H_0$ and model parameters  $\alpha$, $\gamma$, and $\zeta$ constrained by the H(z)+Pantheon+Analysis data sets.

\begin{figure}[H]
\includegraphics[scale=0.3]{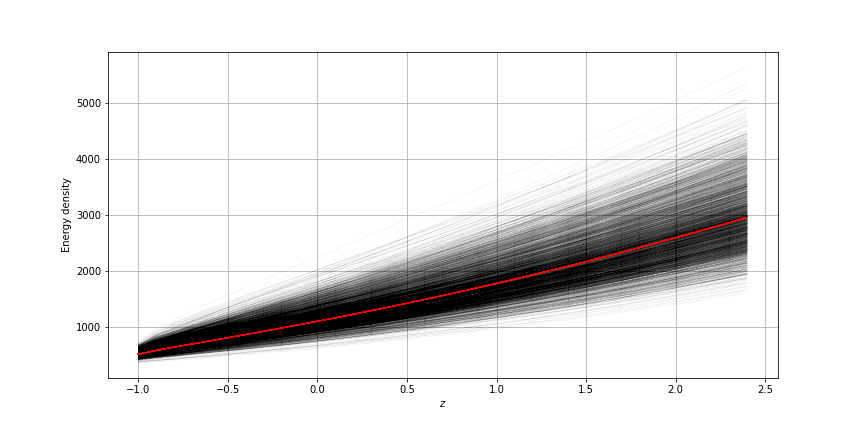}
\caption{The reconstruction of the energy density as a function of the redshift for our model is presented for 7500 samples which are reproduced by re-sampling the chains through \textit{emcee}. We plot all the obtained curves, alongside the curve corresponding to the best fit of the parameters (red curve).}\label{f2}
\end{figure}

\begin{figure}[H]
\includegraphics[scale=0.3]{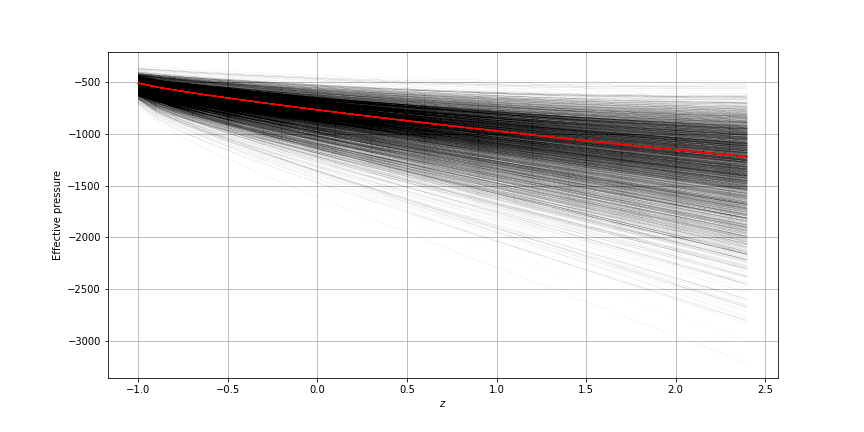}
\caption{The reconstruction of the effective pressure as a function of the redshift for our model is presented for 7500 samples which are reproduced by re-sampling the chains through \textit{emcee}. We plot all the obtained curves, alongside the curve corresponding to the best fit of the parameters (red curve).}\label{f3}
\end{figure}

\begin{figure}[H]
\centering
\includegraphics[scale=0.3]{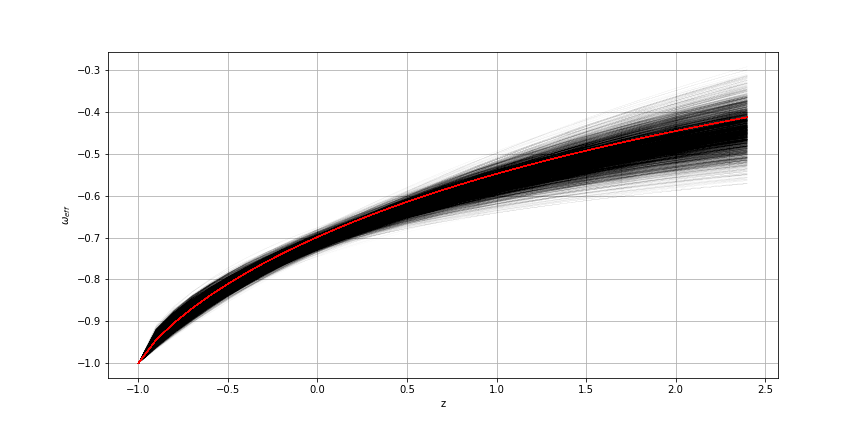}
\caption{The reconstruction of the effective EoS parameter as a function of the redshift for our model is presented for 7500 samples which are reproduced by re-sampling the chains through \textit{emcee}. We plot all the obtained curves, alongside the curve corresponding to the best fit of the parameters (red curve).}\label{f4}
\end{figure}

\justify We reconstructed the matter-energy density, the effective pressure and EoS parameter as a function of the redshift, presented in Figs. \ref{f2}, \ref{f3}, and \ref{f4}, for 7500 samples that are reproduced by re-sampling the chains through \textit{emcee}. From Fig. \ref{f2} it is evident that the cosmic matter-energy density shows expected positive behaviour and it vanishes with the expansion of the universe in the far future. The effective pressure component presented in Fig. \ref{f3} exhibits negative behaviour that can lead to the accelerating expansion of the universe. Further, the present value of the effective EoS parameter is obtained to be $\omega_0 \approx -0.71 $. Thus, the behaviour of the effective EoS parameter in Fig. \ref{f4} confirmed the accelerating nature of the expansion phase of the universe.

\section{Statefinder Diagnostic}\label{sec6}

It is well accepted that the responsible behind cosmic expansion is the dark energy. In the last few decades, the investigation of the origin and fundamental behavior of dark energy is increased. Consequently, plenty of dark energy models started appearing, and therefore the either quantitative or qualitative distinction between these models of dark energy becomes necessary. In this direction, Sahni et al. \cite{V.S.} proposed a statefinder diagnostic method that can identify amongst various dark energy models with the help of a pair of geometrical parameters called statefinder parameters $(r,s)$. It is defined as 

\begin{equation}
 r=\frac{\dddot{a}}{aH^3} 
\end{equation}
and
\begin{equation}
s=\frac{(r-1)}{3(q-\frac{1}{2})}
\end{equation}

\begin{figure}[H]
\includegraphics[scale=0.6]{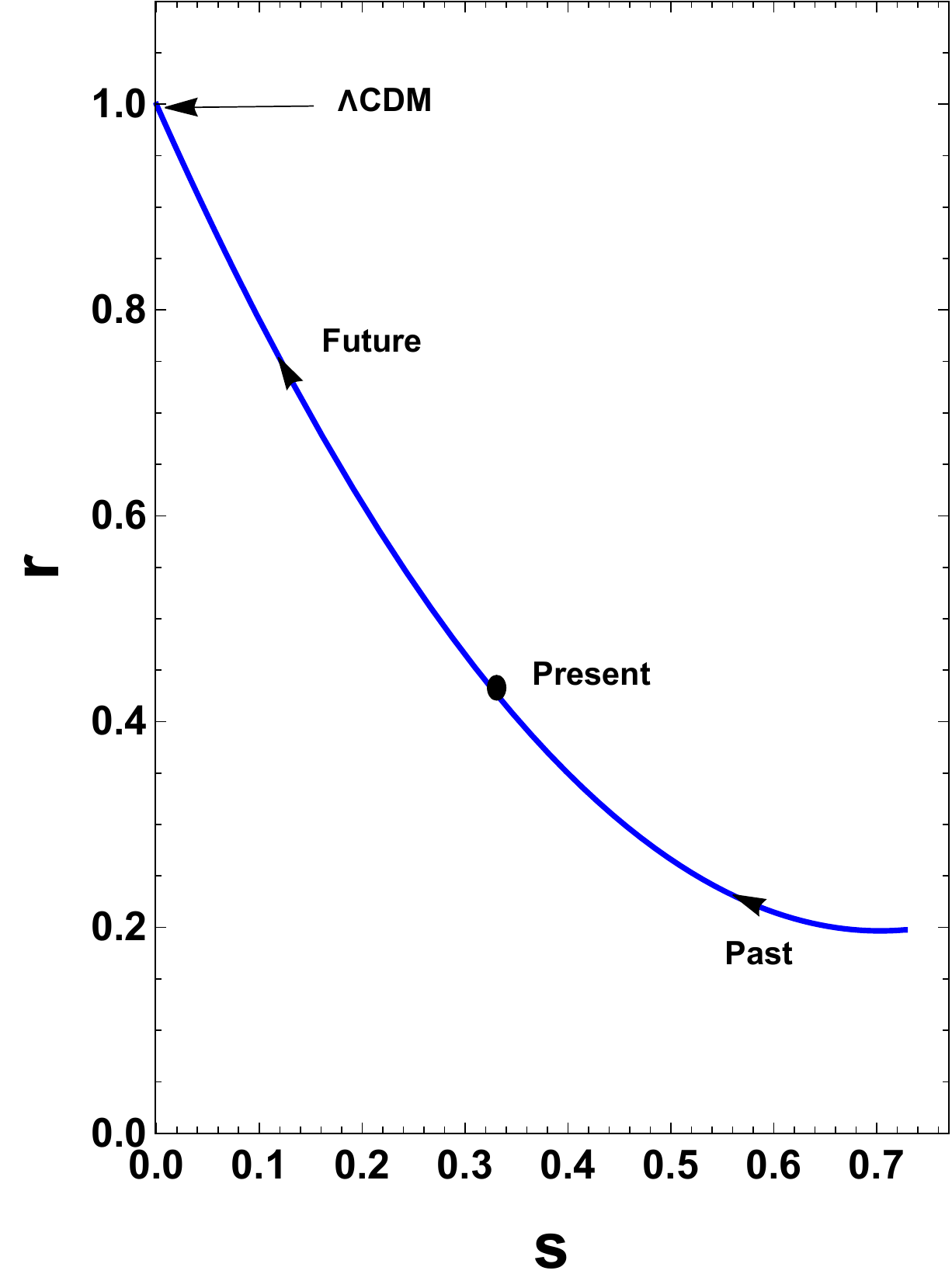}
\caption{Profile of the evolution trajectory of given model in the $r-s$ plane with the agreement of obtained observational constraints.}\label{f5}
\end{figure}

\justify We evaluate the statefinder parameters $(r,s)$ for our cosmological $f(R,L_m)$ model. The evolutionary trajectory of the assumed model with the agreement of obtained observational constraints is presented in Fig. \ref{f5}. The deviation of the evolutionary trajectory of the given model from the $\Lambda$CDM one gives the required discrimination. The value $r=1, s=0$ represents the  $\Lambda$CDM model, $r>1, s<0$ represents the Chaplygin gas model, and $r<1, s>0$ represents the quintessence model. The present value of statefinder parameters for our model is nearly $(r,s)=(0.43,0.33)$. From Fig. \ref{f5} it is evident that the dark component due to modified geometry with the effect of bulk viscosity has quintessence type behavior. 

\section{Om Diagnostics}\label{sec7}

\justify The Om diagnostic is another recently proposed method that can effectively distinguish different dark energy models \cite{Om}. It is much simpler as compared to that of statefinder analysis since it offers the formulation incorporating only the Hubble parameter. For spatially flat constraint, it is given by,

\begin{equation}
Om(z)= \frac{\big(\frac{H(z)}{H_0}\big)^2-1}{(1+z)^3-1}
\end{equation}

\justify The negative slope of $Om(z)$ represents quintessence behaviour whereas positive slope represents phantom behaviour . The constant nature of $Om(z)$ corresponds to the $\Lambda$CDM type behaviour of the given model. 

\begin{figure}[H]
\includegraphics[scale=0.56]{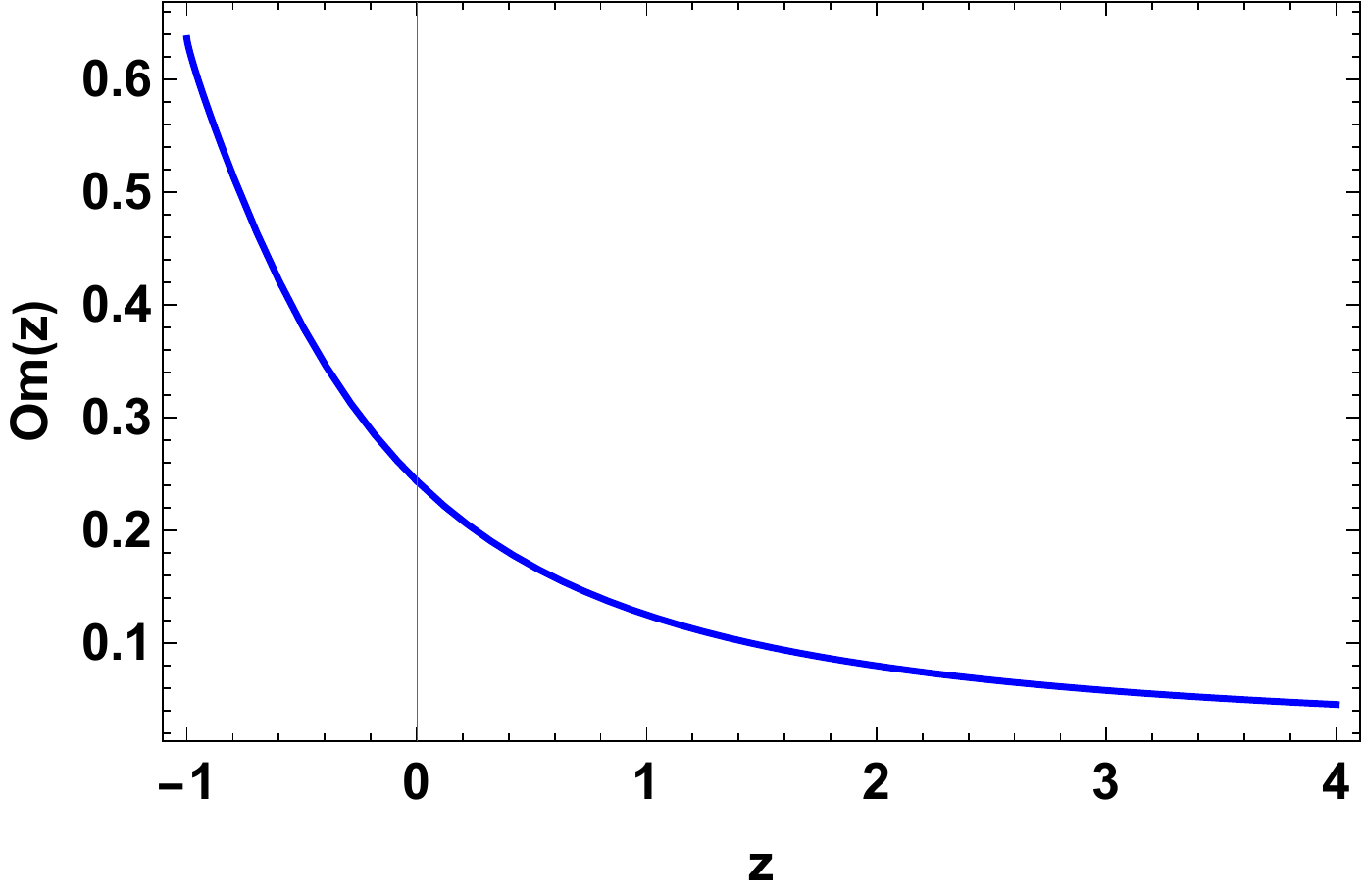}
\caption{Profile of Om diagnostic parameter with the agreement of obtained observational constraints.}\label{f6}
\end{figure}

\justify Fig.\ref{f6} indicate that the Om diagnostic parameter shows negative slope in the entire domain. Thus, Om diagnostic test indicate that our bulk viscous matter dominated $f(R,L_m)$ model follows quintessence scenario.

\section{Energy Conditions}\label{sec8}

Now, we are going to test the viability of the acquired solution corresponding to the assumed $f(R,L_m)$ model by invoking energy conditions criterion. The energy conditions are criteria imposed to the energy-momentum tensor, in order to fulfill the positivity condition of energy. These criteria are offered from the excellent work of Raychaudhuri that is known as Raychaudhuri's equation and are written as \cite{EC} 

\begin{itemize}
\item \textbf{Null energy condition (NEC) :} $\rho_{eff}+p_{eff}\geq 0$;  
\item \textbf{Weak energy condition (WEC) :} $\rho_{eff} \geq 0$ and  $\rho_{eff}+p_{eff}\geq 0$; 
\item \textbf{Dominant energy condition (DEC) :} $\rho_{eff} \pm p_{eff}\geq 0$; 
\item \textbf{Strong energy condition (SEC) :} $\rho_{eff}+ 3p_{eff}\geq 0$,
\end{itemize}
with $\rho_{eff}$ is the effective energy density.

\begin{figure}[H]
\centering
\includegraphics[scale=0.52]{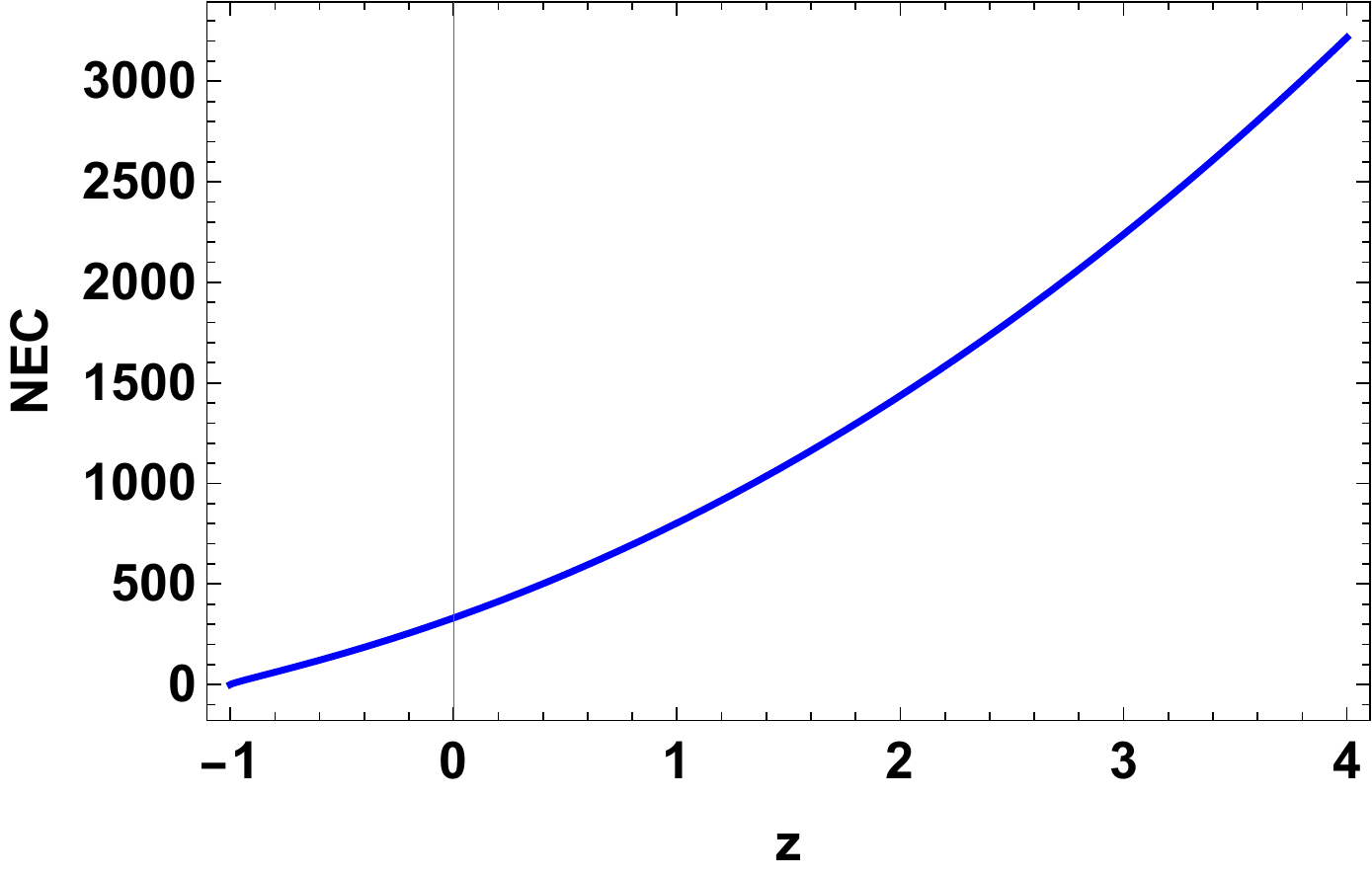}
\caption{Behavior of the NEC vs redshift.}
\label{nec}
\end{figure}

\begin{figure}[H]
\centering
\includegraphics[scale=0.52]{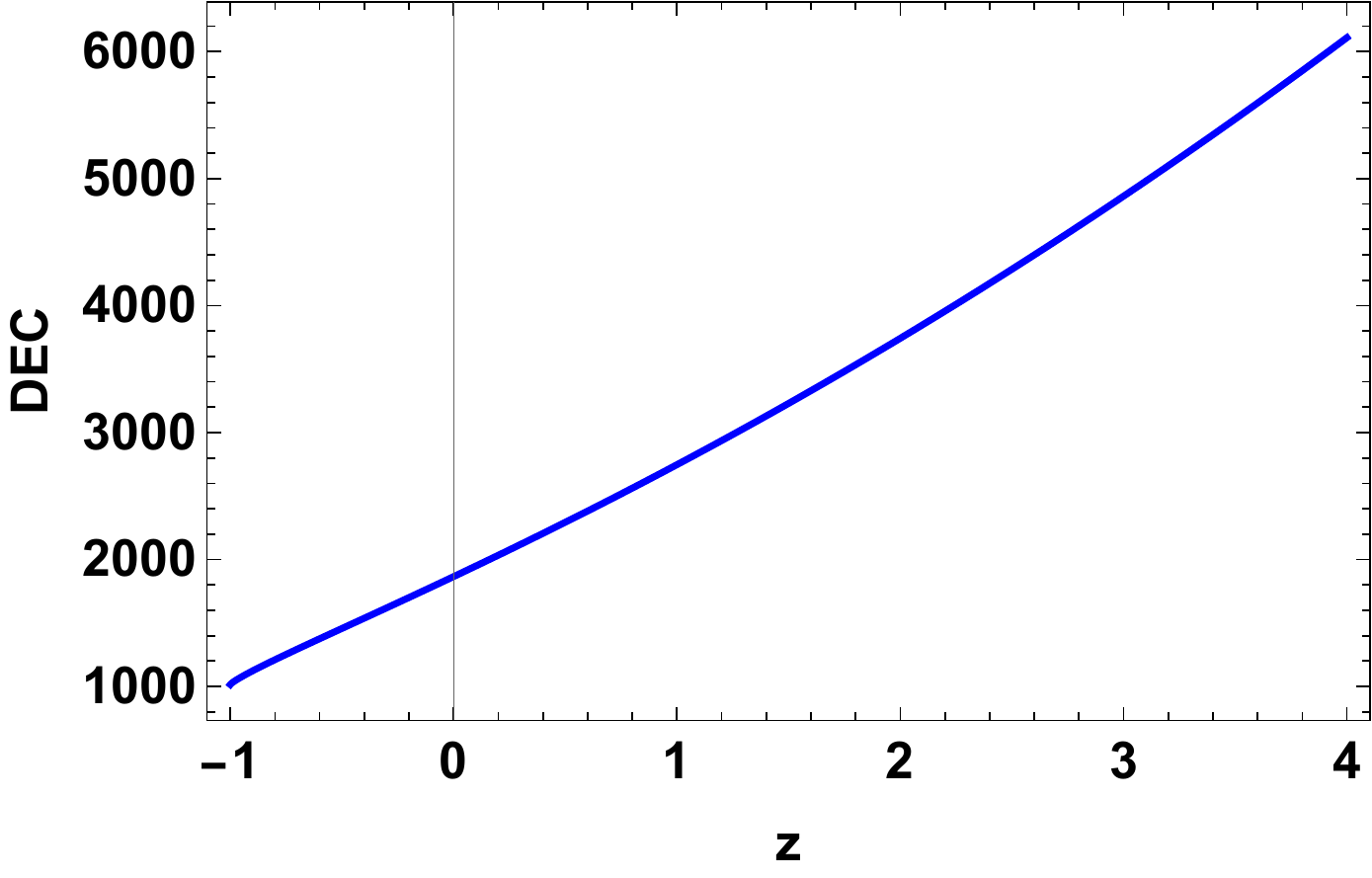}
\caption{Behavior of the DEC vs redshift.}
\label{dec}
\end{figure}

\begin{figure}[H]
\centering
\includegraphics[scale=0.56]{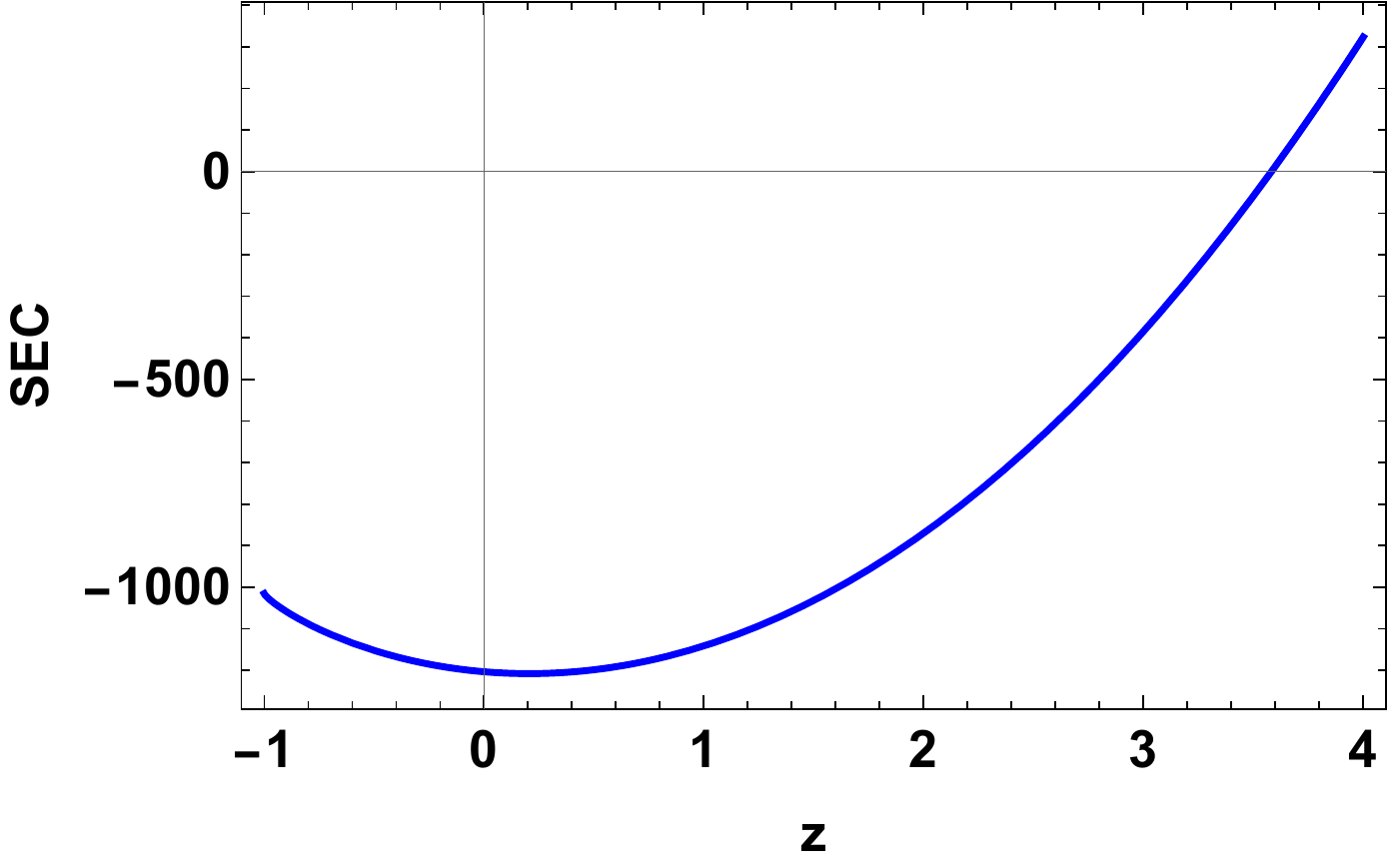}
\caption{Behavior of the SEC vs redshift.}
\label{sec}
\end{figure}

From Figs.\ref{nec} and \ref{dec}, we observed that the NEC and DEC satisfies the positivity criteria in the entire domain of redshift range, corresponding to the estimated values of parameters from observational data sets. As WEC comprises of energy density and NEC, it is also satisfied. Finally Fig.\ref{sec} shows that the violation of SEC occurs in the recent past, and hence this violations favors the cosmic acceleration.

\section{Conclusion}\label{sec9}

\justify Hydrodynamically, the inclusion of the coefficient of viscosity in the cosmic matter content is quite natural, as an ideal characteristics of a fluid is after all an abstraction. In the presented article, we have analyzed the significance of bulk viscosity to drive the cosmic late time acceleration under the $f(R,L_m)$ background. $f(R,L_m)$ gravity theory is a generalization of matter curvature coupling theories \cite{THK-6}. Harko and Lobo investigated the curvature-matter couplings in modified gravity from linear aspects to conformally invariant theories \cite{L-Ex}. Models with non-minimal matter geometry couplings have great astrophysical and cosmological implications \cite{THK-4,THK-5,THK-EX,V.F.-2}. For our analysis, we considered a $f(R,L_m)$ function, particularly, $f(R,L_m)=\frac{R}{2}+L_m^\alpha $, where $\alpha$ is free model parameter. Then we have assumed the effective equation of state in equation \ref{2e}, which is the Einstein case value with proportionality constant $\zeta$ used in the Einstein theory \cite{IB-1} frequently used in the literature. We found the exact solution of our bulk viscous matter dominated $f(R,L_m)$ model, and then we used the combined $H(z)+Pantheon+Analysis$ observational data sets to constrain the present value of the Hubble parameter $H_0$ and the model parameters. The obtained best fit values are $\alpha=1.310^{+0.037}_{-0.032}$, $\gamma = 1.29 \pm 0.20$, $\zeta= 5.02 \pm 0.26$, and $H_0= 72.09 \pm 0.19$. In addition, we have characterized the behaviour of matter-energy density, pressure component incorporating viscosity, and the effective EoS parameter as a function of the redshift, presented in Figs. \ref{f2}, \ref{f3}, and \ref{f4}, for 7500 samples that are reproduced by re-sampling the chains through \textit{emcee}. From Fig. \ref{f2} it is evident that the cosmic matter-energy density shows expected positive behaviour and the effective pressure component presented in Fig. \ref{f3} exhibits negative behaviour that can lead to the accelerating expansion of the universe. Moreover, the present value of the effective EoS parameter is obtained to be $\omega_0 \approx -0.71 $. Thus, the trajectory of the EoS parameter in Fig. \ref{f4} confirmed the accelerating nature of the expansion phase of the universe. Then we evaluated the $(r,s)$ parameters for our assumed $f(R,L_m)$ model. The present value of statefinder parameters for our model is nearly $(r,s)=(0.43,0.33)$. From Fig. \ref{f5} we noticed that the evolutionary trajectory of our $f(R,L_m)$ model lies in the quintessence region. Further, the Om diagnostic presented in Fig \ref{f6}, indicates that our assumed $f(R,L_m)$ model favors the quintessence type dark energy. Finally, the energy conditions presented in Fig \ref{nec}, \ref{dec}, and \ref{sec}, exhibits positivity criteria in the entire domain of redshift range corresponding to the case of NEC and DEC, whereas it shows the violation in case of SEC. This violation of SEC, occurs in the recent past, favors the observed acceleration. We conclude that our cosmological $f(R,L_m)$ model with the fluid incorporating the bulk viscosity effects, can efficiently interpret the late time cosmic phenomenon of the universe with observational compatibility.

\section*{Data Availability Statement}
There are no new data associated with this article.

\section*{Acknowledgments} \label{sec11}
L.V.J. acknowledges University Grant Commission (UGC), Govt. of India, New Delhi, for awarding JRF (NTA Ref. No.: 191620024300). R.S. acknowledges UGC, New Delhi, India for providing Junior Research Fellowship (UGC-Ref. No.: 191620096030). SM acknowledges Department of Science and Technology (DST), Govt. of India, New Delhi, for awarding Senior Research Fellowship (File No. DST/INSPIRE Fellowship/2018/IF18D676). We are very much grateful to the honorable referees and to the editor for the illuminating suggestions that have significantly improved our work in terms of research quality, and presentation.


\end{document}